\documentclass[12pt]{iopart}

\usepackage{graphicx}
\usepackage{epsfig}

\begin{document}
\title[Numerical simulation of long biopolymer translocation]{Numerical simulation of conformational variability in biopolymer
translocation through wide nanopores
}

\author{Maria Fyta$^1$\footnote{Present address: Department Physik, Technische Universit$\ddot{a}$t M$\ddot{u}nchen$,
Garching, 85747 Germany}, Simone Melchionna$^{2,3}$, Massimo Bernaschi$^4$,
Efthimios Kaxiras$^{1,2}$, and Sauro Succi$^{4,5}$}

\address{$^1$ Department of Physics and $^2$ School of Engineering and Applied
Sciences, Harvard University, Cambridge, MA, USA\\
$^3$ INFM-SOFT, Department of Physics, Universit\`a di Roma 
\it{La Sapienza}, P.le A. Moro 2, 00185 Rome, Italy \\
$^4$ Istituto Applicazioni Calcolo, CNR, 
Viale del Policlinico 13, 00161, Roma, Italy \\
$^5$ Initiative in Innovative Computing, Harvard University,
Cambridge, MA, USA\\
}

\ead{mfyta@seas.harvard.edu}
\date{\today}

\begin{abstract}
Numerical results on the translocation of long biopolymers through mid-sized and wide pores
are presented. The simulations are  based on a novel methodology which couples molecular motion 
to a mesoscopic fluid solvent. Thousands of events of long polymers (up to 8000 monomers) are monitored
as they pass through nanopores.
Comparison between the different pore sizes shows that wide pores can
host a larger number of multiple biopolymer segments, as compared to smaller pores.
The simulations provide clear evidence of folding quantization in the translocation
process as the biopolymers undertake multi-folded configurations, characterized by a
well-defined integer number of folds.
Accordingly, the translocation time is no longer represented by a single-exponent power law dependence on the length, as
it is the case for single-file translocation through narrow pores.
The folding quantization increases with the biopolymer length, while
the rate of translocated beads at each time step  is linearly 
correlated to the number of resident beads in the pore. 
Finally, analysis of the statistics over the translocation work unravels the importance
of the hydrodynamic interactions in the process.
\end{abstract}


\maketitle

\section{Introduction}

Biological systems exhibit a complexity and diversity far richer than the simple solid or fluid systems traditionally studied in physics or chemistry. Advances in computer technology and breakthroughs in computational methods have been constantly reducing the gap between quantitative models and actual biological behavior. The main challenge remains the wide range of spatio-temporal scales involved in the dynamical evolution of complex biological systems. In response to this challenge, various strategies have been developed recently, based on composite computational schemes in which information is exchanged between the scales.
Motivated by recent experimental studies \cite{Li_NatMat2003,Storm_NanoLett2005}, we apply such a computational scheme to the translocation of a biopolymer through nanopores. 
The translocation of biopolymers plays a major role in many important biological processes, such as viral infection by phages, inter-bacterial DNA transduction, and gene therapy \cite{TRANSL}. 
The ultimate goal of these studies is to open a path  for ultra-fast DNA-sequencing by sensing the base-sensitive electronic signal as the biopolymer passes through a nanopore with attached electrodes. The importance of this process has spawned a number of in vitro experiments, aimed at exploring the translocation process through micro-fabricated channels \cite{deamer07} under the effects of an external electric field, or through protein channels across cellular membranes \cite{EXPRM1,EXPRM2}. From a theoretical point of view, simplified schemes \cite{KARDAR,statisTrans}, coarse-grained or microscopic models with and without hydrodynamic interactions \cite{DynamPRL,Forrey07,Nelson} or mesoscopic approaches \cite{reboux06} are able to analyze universal features of the translocation process. However, a quantitative description of this complex process, 
which involves the competition between many-body interactions at the atomic or molecular scale, fluid-atom hydrodynamic coupling, as well as the interaction of the biopolymer with wall molecules in the region of the pore, calls for state-of-the art modeling, towards which the results presented here are directed.

In a recent paper, the translocation of biopolymers through (relatively) 
large pores was reported to exhibit the intriguing phenomenon of 
current-blockade quantization \cite{Li_NatMat2003}. 
This was interpreted as an {\it indirect} evidence that the polymer crosses 
the pore in the form of "quantized" configurations, associated with 
integer values of the folding number, the number of strands simultaneously 
occupying the pore during the translocation.
Such a behavior has been recently confirmed by the {\it direct}
observation of multi-folded configurations in large-scale numerical 
simulations of biopolymer translocation by the present authors \cite{ourNL}.
Here, we considerably extend the scope of such simulations, by increasing the
polymer lengths up to $8000$ monomers,
about an order of magnitude above any previous simulation in the field. This unexplored regime of polymer 
lengths reveals an extremely rich configurational dynamics, especially 
in larger pores. In this work, we elaborate more on this enriched dynamics
and analyze in detail the phenomenon of folding quantization.

\section{Multiscale scheme}

Our simulations are based on a multiscale methodology \cite{ourLBM}, which involves 
the coupling of a mesoscopic lattice Boltzmann (LB) \cite{LBE} 
approach for the solvent degrees of freedom and molecular dynamics (MD)
for the biopolymer motion.
The comparison of our previous results to those of experiments of DNA translocation,
allows us to map the anonymous simulated polymers to actual biopolymers \cite{ourPRE}.
A three-dimensional box of size $(N_x  \times N_y  \times N_z ) \Delta x$ 
lattice units, with $ \Delta x$ the spacing between lattice points
surrounds both the solvent and the polymer.
We take $N_x = 2 N_y$, $N_y = N_z$ and $N_x = 128$ and 
biopolymers with $N_0 = 100 - 8000$ beads, spanning
nearly {\it two} orders of magnitude in polymer length.
At $t=0$ the polymer resides entirely in the right chamber 
at $x>\frac{N_x}{2}\Delta x$. 
A separating wall is located in 
the mid-section of the $x$ direction, at $x=\frac{N_x}{2}\Delta x $.  
At the center of this wall, 
a cylindrical pore of length $l_{p}=3 \Delta x$ and diameter $d_{p}$ is opened up.
We have used two different pore diameters, 
a small $d_p=5$ and wide $d_p=9$ in units of $\Delta x$.
Translocation is induced by a constant electric field, localized around the pore, similarly 
to the experimental settings \cite{Storm_NanoLett2005},
acting along the $x$ direction and confined to a cylindrical
channel of the same size as the pore and length $3\Delta x$ along the 
streamwise ($x$) direction. 
All parameters are measured in units of the LB
time step and spacing, $ \Delta t$ and  $\Delta x$, respectively,
which are both set equal to $1$.
The MD time step is $5$ times smaller than $ \Delta t$.
With the pulling force associated with the electric field
in the experiments set at $0.02$ and the temperature at $10^{-4}$, the process falls in the  {\it fast} translocation regime.

The monomers interact through a Lennard-Jones 6-12 potential with parameters
$\sigma=1.8 $, $\epsilon=10^{-4}$ and a cut-off at $2.02$. The interaction of
the monomers with the wall is modeled by a Lennard-Jones 6-12 potential with parameters
$\sigma_w=1.5 $, $\epsilon_w=10^{-3}$ and a cut-off at $1.68$.
Accordingly, the effective width and radius of the surrounding pore should take into
account the repulsive monomer-wall interactions, so that a monomer is counted 
as being inside the pore if contained in a cylinder of effective width $\simeq 6 \Delta x$ and radius 
$\simeq 3.5$ and $7.5$ for $d_p=5$ and $9$, respectively.
The bonds between adjacent beads are modeled through springs, with a
spring constant $k=0.5$ and an effective equilibrium 
bond length  $b=1.2$.  The solvent has density $\rho=1$,
kinematic viscosity $\nu=0.1$ and damping coefficient with the embedded particles $\gamma=0.1$. 
Further details can be found in Ref. \cite{ourLBM}.
One lattice spacing $\Delta x=42$ $nm$, so that
the bond length $b=1.2 \; \Delta x$ corresponds to the
persistence length of double-stranded DNA ($50nm$). 
In measuring the residence number of beads in the pore region we have defined a
cylinder of length $10\,\Delta x$ and radius $d_p$ centered at the pore midpoint 
and with its axis aligned with the pore axis. 

\section{Configurational analysis}

\begin{figure}
 \begin{center}
 \includegraphics[width=1.0\textwidth]{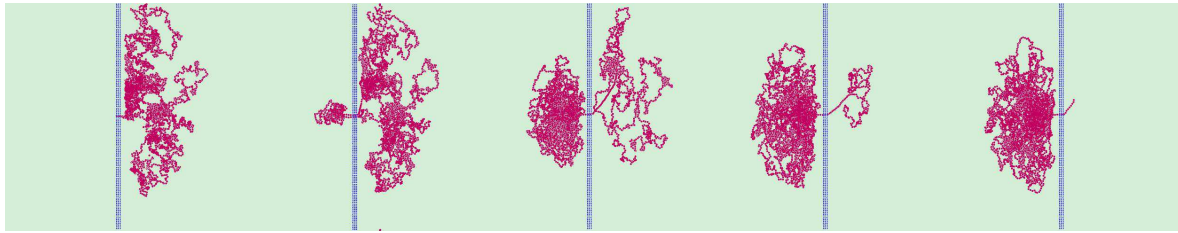}
 \caption{Snapshots of a biopolymer ($N_0=4000$) translocating
through a wide pore ($d_p=9$). Different folding conformations
are visible. A force (not shown) is applied
at the pore region and initiates translocation from right to left.
In order to reveal the polymer
conformation as it passes though the pore, the wall is shown thinner than it 
actually is in the simulations.
 \label{fig:snaps}}
 \end{center}
 \end{figure}

The ensemble of simulations is generated by different realizations
of the initial polymer configuration, to account for the 
statistical nature of the process.
The combined statistics over initial conditions and time evolution of the
simulations, delivers an aggregate 
ensemble ranging from $500,000$ events for the shortest history ($N_0=100$, $d_p=9$) up
to nearly 10 million events for the longest one ($N_0=8000$, $d_p=5$).
Fig. \ref{fig:snaps} shows different snapshots
of such a translocation event for a $N_0=4000$ long biopolymer.
In the initial stage of the translocation, the nanopore
gets populated, with the biopolymer undertaking a high-fold conformation
as it passes through the pore.
The range of the number of folds explored by the translocation
trajectories grows approximately with the cross-section of the pore
and the polymer length.

\begin{figure}
\begin{center}
\includegraphics[width=0.75\textwidth]{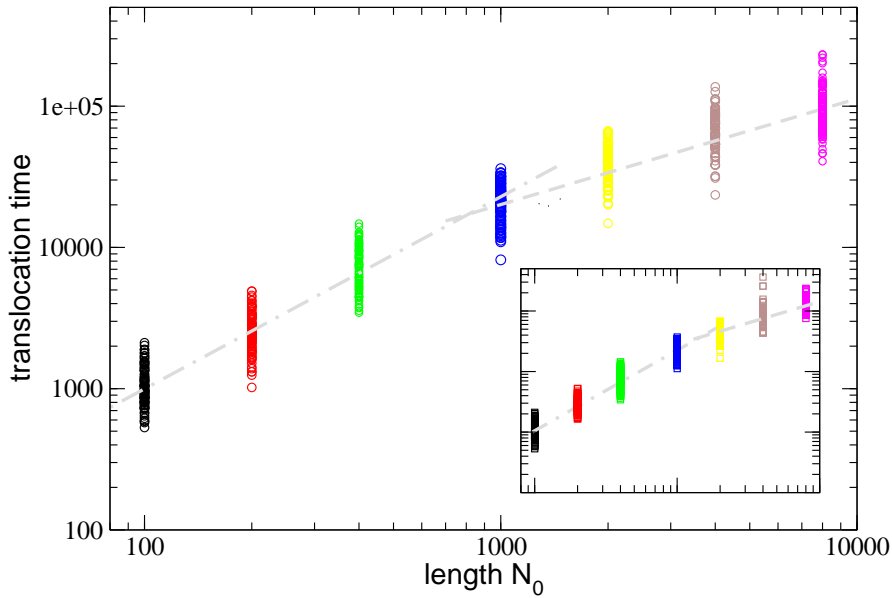}
\caption{Scatter plots of all events for biopolymers translocating through the wide pore ($d_p=9$).
The two different exponents, $1.36$ and $0.75$ (see text) are denoted by the two lines. 
The inset shows similar events through a narrower pore ($d_p=5$).
Again, the two lines imply the two exponents $1.31$ and $0.70$.}
\label{fig:Translotime}
\end{center}
\end{figure}

The observed translocation time results from a weighted average over a whole set of multi-folded configurations 
attained by the polymer during translocation,
The weights in this average depend on the number of folds a polymer of
a given length undertakes within the pore, and correspond to the probability
distribution function (number of counts normalized to unity) shown in Fig. \ref{fig:qprob}.
Within this picture, a single scaling exponent characterizing 
the translocation time as a function of the polymer length, appears to be 
insufficient because, above a given length, many folded-states are simultaneously 
excited, as their number is an increasing function of the pore diameter. 
The translocation time of these multi-folded configurations is shown to be dominated by the
low-folded states, which explains the relatively minor deviations of 
the translocating time from the single-file power-law expression 
$\tau \sim N_{0}^{1.27}$ \cite{Storm_NanoLett2005}, where $\tau$ and $N_0$ 
are the translocation time and the length of the biopolymer, respectively. 
In order to visualize this behavior, in Fig. \ref{fig:Translotime}, we report the translocation time as a 
function of the polymer length, $N_0$, for both pores and all events for each length. 
The figure shows that, up to a length of $N_0=1000$ for the
wide ($d_p=9$) and $N_0=2000$  for the narrower ($d_p=5$) pore, the most probable translocation time
$\tau$ obeys a scaling law of the form $\tau \sim N_{0}^{\alpha}$, with 
$\alpha \sim 1.31$ and $\alpha \sim 1.36$, respectively. These values
are slightly larger than the corresponding values for
narrow nanopores \cite{Storm_NanoLett2005,ourLBM}. 

This feature does not occur only due to statistical uncertainities, but may also depend on the
pore width. Polymers translocating across a pore with $q>1$, where $q$ denotes
a $q$-folded configuration, possibly enhance the resistance to the electric drive, due to the additional energy 
needed to keep them folded against internal flexional-relaxation forces.
These forces tend to un-fold the polymer, increasing
its interaction with the walls, which may slow down the process as compared
with the single-strand translocation \cite{Forrey07,Nelson,POLPOR}.
By restricting the analysis to the longest 
chains, $N_0\geq 1000$ ($N_0\geq 2000$) for the wide (narrow) pore,  the 
bending of the curve could be interpreted 
as the emergence of a new scaling exponent, $\alpha_2 \sim 0.75$ (0.70). 
Evidently, there are not enough data points to support the exact values for
the scaling exponents, but the significant decrease of these exponents for
long biopolymers is qualitatively apparent.
In Fig. \ref{fig:Translotime}, the two different exponents are denoted by
the grey lines (with a different prefactor in each case in order to match
the scaling law to the simulation data).
We have found that this bending is due to the multi-fold 
conformation of the translocating biopolymer, which does not necessarily 
follow a power-law dependence on the polymer size. 
As a result, in contrast to single-file translocation (the case for narrow pores), 
multi-fold translocation does not need to obey a standard power-law scaling. 


\begin{figure}
 \begin{center}
 \includegraphics[width=0.7\textwidth]{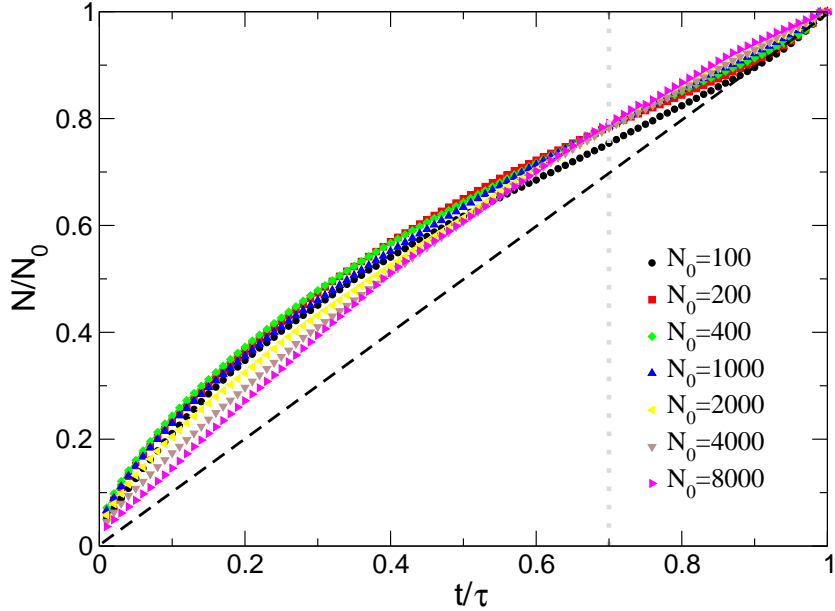}
 \caption{Scaled number of translocated beads with time for all
lengths ($N_0=100-8000$). The pore diameter is $d_p=9$ and both
axes are scaled with respect to the total number of beads and
the total translocation time, respectively. The black dashed line
corresponds to a constant translocation-speed, $dN(t)/dt=N_0/\tau$.
The vertical dotted line shows the crossover discussed in the text.
 \label{fig:ntra}}
 \end{center}
 \end{figure}

An insightful way to monitor the translocation process is through the evaluation
of the fraction of translocating beads with time. Averages
over all events for all lengths are plotted in Fig. \ref{fig:ntra},
where both the number of the translocated beads and the time are
presented in reduced units, i.e. scaled with respect to the total number of beads and the total
translocation time for each event and length. 
In such units, universality would result in a collapse of the translocation data
at all lenghts into a single master curve. 

Fig. \ref{fig:ntra} shows a clear trend with size, which only
very short chains ($N_0=100$) do not follow. 
Initially, the shorter biopolymers exhibit a larger translocation speed and also a larger acceleration, 
but at the final stages this trend is reversed, as the longer chains accelerate and
eventually translocate faster through the pore. The crossover, where the
translocation speed of the long biopolymers  becomes larger than the corresponding of the
shorter ones occurs at $0.7N_0$ (see dotted line in Fig.\ref{fig:ntra}),
a point at which $70$\% of the chain has already translocated.
It is interesting to observe that this is a universal value for 
all lengths studied here, although at this point, this is a purely
observational fact. Again, only polymers with
$N_0=100$ do not follow this trend, as these are too short.
For all times, the beads follow a super-linear trajectory compared
to the constant speed translocation (dashed line in the figure),
but at the final stages the end part of the polymers translocate
with constant speed.
In the case of the narrower pore ($d_p=5$), the trend is similar, although
no universal crossover close to the constant speed limit is observed.

\subsection{Quantization of the folding number}

\begin{figure}
\begin{center}
\includegraphics[scale=0.55]{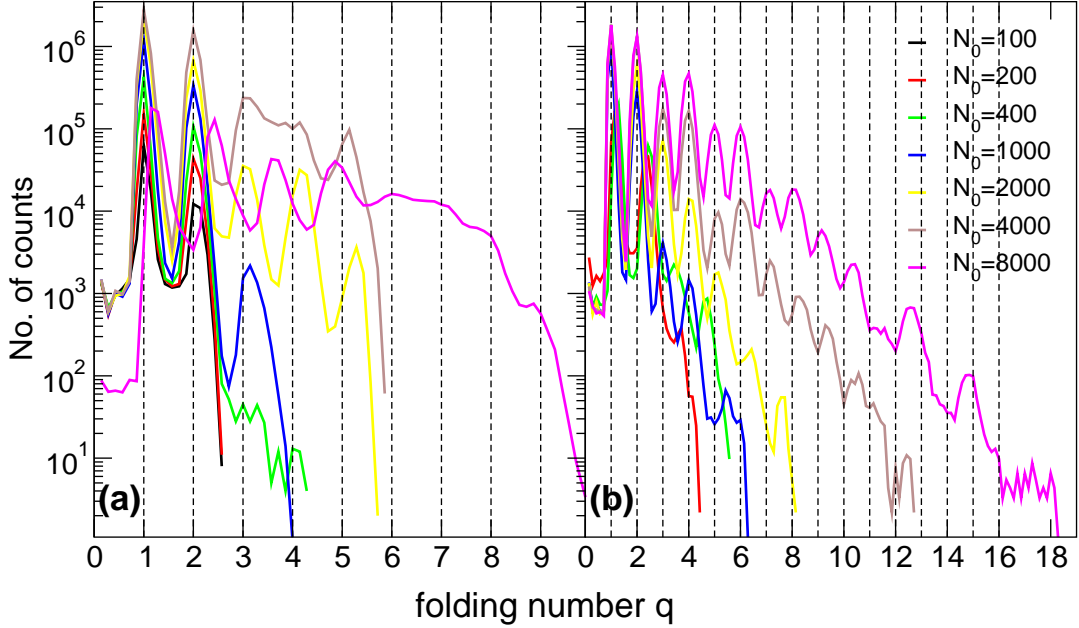}
\caption{Probability distribution of the folding number $q$ for the entire set of polymer
lengths, $N_0=100-8000$, and both pore diameters (a) $d_p=5$ and (b) $d_p=9$.}
\label{fig:qprob}
\end{center}
\end{figure}

In order to investigate the quantization of the resident beads in the pore,
we monitored the distribution of the number of pore-resident beads 
$N_r$ with the number $q$ for a $q$-folded translocation. 
The resident monomers block the current across the channel, so that 
$N_r$ conveys a direct measure of the
current drop associated with the biopolymer passage through the nanopore.
All distributions are peaked around quantized values
of $q$ (as defined below). A large fraction of the events are around $q=1$, hence single-file
translocation, which is also the conformation at the late stages of
the process for all lengths.
In Fig. \ref{fig:qprob}, the cumulative statistics of the 
folding number, $q=N_r/N_{1}$, are shown as collected at each time-step of every 
single trajectory for a series of $100$ realizations for each polymer length.
Here, $N_{1}\sim ~7$ is the single-file value of the resident number, 
as also confirmed by visual inspection of the configurations. 

The wide pore ($d_p=9$) reveals a sharp quantization of the distribution of the
folding number, with a very well defined peak in the distribution.
Simulation data show, that the range of $q$-numbers, occupied in this case grows up to $q\sim 10$.
The average $q$ over all realizations for each length remains approximately
constant $\sim~1.3$, up to $N_0 < 1000$, and grows up to about $2.6$ 
for $N_0=8000$.
Accordingly, the departure of the translocation
time from a power-law at large $N_0$ is mainly due to the increase of 
the average $q$ with the polymer length (for $N_0 >1000$). 
This results from the shift of the probability distribution of the translocation time 
towards higher $q$-folds as $N_0$ increases.
For all lengths inspected, the time average of the folding number $q $ remains below
$3$, because the states $q=1$ and $q=2$ continue to be the most populated ones.

\begin{figure}
\begin{center}
\includegraphics[width=0.65\textwidth]{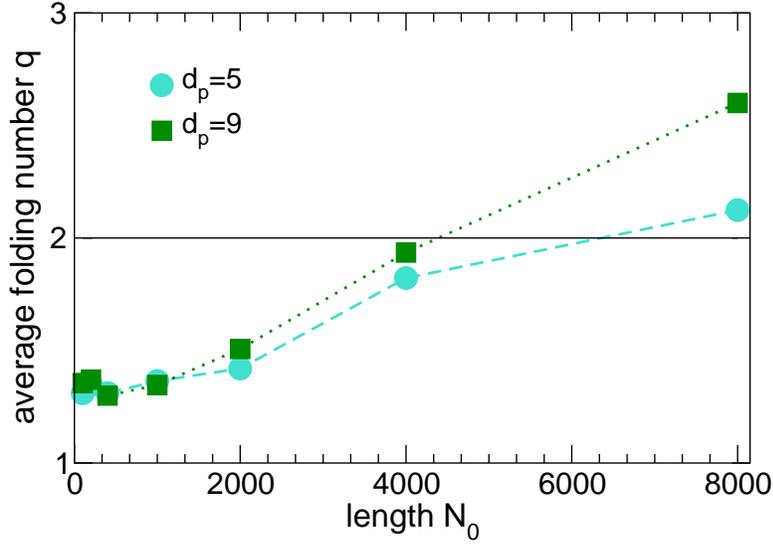}
\caption{The average folding number $\bar q$ as a function of the polymer length for the wide ($d_p=9$) and 
narrower ($d_p=5$) pore, respectively.}
\label{fig:aveq}
\end{center}
\end{figure}

The narrower pore ($d_p=5$) shows the same trends, although on a smaller range
of $q$-numbers, up to $q\sim 6$. Above this value, the quantized peaked structure
in the folding probability is lost.
For this pore, the average $q$ over all realizations for each length remains close
to $1.3$, up to $N_0 < 2000$, and grows up to about $2.1$ for $N_0=8000$.
These data indicate that quantization of the folding number is better manifested
by long chains crossing wide pores.
The fact that the average $q$ remains the same ($\sim~1.3$) for both pores and relatively short
biopolymers shows that, for these lengths, there is no essential effect by varying the pore size.
On the contrary, for a wide pore, the average $q$ increases faster 
in the range of long biopolymers than for a narrower pore.
It is quite natural to expect that long polymers transiting
through large pores prove capable of supporting a widely richer spectrum of folded conformations
as compared to the case of short polymers transiting through narrow pores.
Graphical evidence for these  points is given in Fig. \ref{fig:aveq}, where
the average folding number $q$ for all cases studied is plotted as a function of
the length $N_0$. Evidently, the average $q$ is almost constant up to $N_0=1000-2000$
and increases thereafter, exposing the difference between the two 
pore sizes, which remains nonetheless rather mild.
The highest-folding number supported by the polymer-pore system should
increase quadratically with the pore diameter, as the number of monomers
a given pore can accomodate is clearly proportional to the 
cross section of the pore.

\section{Forces influencing the translocation process}

The translocation dynamics depends on the strength of the frictional forces exerted by the wall.
In single-file translocation, it is well known that strong friction changes the power-law exponent from
$\simeq 1.2$ to a linear dependence $\tau \propto N_0$ \cite{diventra}. 
In the case of multi-file translocation, it is
unclear whether the high-fold configurations induce high-friction conditions, that
would decrease the rate of the translocated beads ($\dot N=\frac{dN}{dt}$) with the resident
monomers ($N_{r}$) for high values of the folding number $q$. We have 
observed that $\dot N$ is linearly correlated to $N_{r}$ with basically the same average slope for all folds. 
Therefore, frictional forces affect a small layer
close to the wall, and have little effect on the group of monomers 
translocating in the inner region of the pore.
This seems to rule out the possibility that the change of exponent be caused by the pore frictional forces. 
Supporting data from about 100 events
for the longest biopolymer ($N_0=8000$) translocating through the wide pore ($d_p=9$) reveal that
all events are centered around integer values of the folding number $q$. As the characteristic 
number $q$ increases, the rate of the translocating monomers with time increases linearly.
Inspection of the results for all lengths
($N_0=100-8000$) and both pore lengths ($d_p=9$ and $d_p=5$), shows that (a) for the same pore, 
the slope of the $q$-$\dot{N}$ curve increases with increasing chain length 
(long polymers slightly slow-down), and (b) for the same polymer length, the linear correlation 
between  $q$ and $\dot{N}$ is better manifested for wider pores.

\begin{figure}
 \begin{center}
 \includegraphics[width=0.75\textwidth]{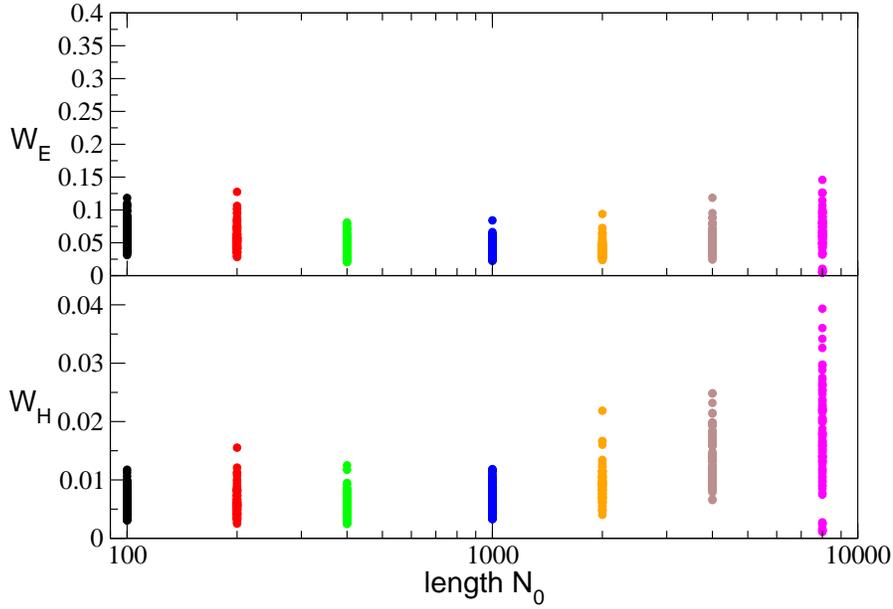}
 \caption{Scatter plot of the translocation work: (top panel)
the work for the driving field and (low panel) the work for the
hydrodynamic field as functions of length $N_0$ for translocation
through the wide pore ($d_p=9$).
 \label{fig:work}}
 \end{center}
 \end{figure}

Another set of forces, which greatly affect the translocation are the fluid-biopolymer interactions
and the driving force at the pore region.
The simulations reveal that solvent {\it correlated}
motion makes a substantial contribution to the translocation energetics.
The role of hydrodynamic correlations is best highlighted by computing the work
done by the moving fluid on the polymer, $W_{H}$, over the entire translocation
process as compared to the case of a passive fluid at rest. 
Inspection of all events for all polymers translocating through wide pores, shown
in Fig. \ref{fig:work}, reveals that the cooperation of the surrounding solvent and the solute
monomers is larger as the length $N_0$ of the polymer increases. 
The work of the driving force, $W_E$, (always positive), varies only slightly
with the length $N_0$, as compared to the corresponding behavior of the
hydrodynamic work. A rough estimate shows that, as the length increases from
$N_0=100$  to $N_0=8000$, there is an increase of about $150\%$ in $W_H$, while 
the corresponding increase in $W_E$ is less than $10\%$. (For these estimates, we
used the average values of $W_H$ and $W_E$ for each length $N_0$).
For a narrower pore ($d_p=5$), these effects are still visible, though
not as strong revealing a higher cooperativity of the solvent 
during translocation through wide pores.

\section{Summary}

In summary, multi-scale simulations of the translocation of 
long polymers, consisting of up to $8000$ beads, across wide nanopores,  
capable of hosting multi-file configurations, provide clear evidence of a 
sharp quantization of the translocation process. 
Throughout their translocation, the polymers undertake 
multi-folded configurations, associated with well-defined integer folding numbers.
The observed translocation time reveals a bending of the scaling law with the
biopolymer length, which gives rise to two different exponents, one of them
describing the short biopolymers and the other the longer ones.
The longer the polymer and wider the pore, the quantization of conformational folds
is more evident and is accompanied by an enhancement of the synergistic role
of the hydrodynamic field.

\ack
SM and MB acknowledge support by the Initiative in Innovative Computing
at Harvard University.

\end{document}